\RequirePackage{fixltx2e}
\documentclass[a4paper]{isp}
\usepackage[american]{babel}
\usepackage{graphicx}
\usepackage{amsmath,amsthm,mathtools}
\usepackage{url}
\usepackage[capitalise,nameinlink]{cleveref}
\usepackage{xcolor}

\def\jcap{Journal of Cosmology and Astroparticle Physic  } 
 
\def\procspie{Proceedings of the SPIE}%

\begin{document}
\pdfgentounicode=1
\title
{
New probability distributions in astrophysics:
III. The truncated Maxwell-Boltzmann distribution
}
\author{Lorenzo  Zaninetti}
\institute{
Physics Department,
 via P.Giuria 1, I-10125 Turin,Italy \\
 \email{zaninetti@ph.unito.it}
}

\maketitle

\begin {abstract}
The Maxwell-Boltzmann (MB) distribution 
for velocities in ideal gases is usually 
defined between zero and infinity.
A double truncated  MB distribution is here 
introduced  and 
the probability density function, 
the distribution function, 
the average value,
the rth moment about the origin,
the root-mean-square speed
and the variance are evaluated.
Two applications are presented:
(i) 
a numerical relationship between 
root-mean-square speed and temperature, and
(ii) 
a modification of the formula for  the Jeans escape flux
of molecules from an atmosphere.
\end{abstract}

\section{Introduction}

The {\em Maxwell-Boltzmann} (MB) distribution, see
\cite{Maxwell1860,Boltzmann1872},
is a powerful tool to explain 
the kinetic theory of gases.
The range in velocity of  this distribution spans
the interval $[0, \infty]$, which produces 
several problems
\begin{enumerate}
\item
The maximum velocity of a gas cannot
be greater than the velocity of light, c  
\item 
The kinetic theory is developed in  a classical
environment, which means that the
involved velocities should be smaller than 
$\approx\,1/10\,c$
\end{enumerate}
These items point toward the hypothesis 
of an upper bound in velocity for the MB.
We will now report some approaches,  
including  an  upper  bound in velocity:
the ion velocities parallel
to the magnetic field in a low density 
surface of a ionized plasma \cite{Buzzi1970};
propagation of longitudinal electron waves in a collisionless, homogeneous,
isotropic plasma, whose velocity distribution function is a truncated
MB \cite{Treguier1976};
fast ion production in laser plasma \cite{Kishimoto1983};
the  release of a dust particle from a 
plasma-facing wall \cite{Tomita2007}; 
an explanation of an anomaly 
in the Dark Matter (DAMA) experiment  \cite{Fowlie2017};
a distorted MB  distribution of epithermal ions 
observed associated with the collapse of energetic ions
\cite{Ida2017};
and
deviations to MB distribution that could have observable effects
which can be measured trough the vapor spectroscopy at an interface
\cite{Todorov2019}.
However, these approaches  do not clearly cover  
the effect of introducing a lower and an upper boundary 
in the MB distribution, which is the subject 
that will be analyzed in this paper.

This paper is structured as follows. 
Section \ref{section_mb}
reviews the basic statistics of the MB distribution
and it derives a new approximate expression for the 
median.
Section \ref{section_mb_truncated}  introduces 
the double truncated MB and it derives  the connected statistics.
Section \ref{section_mb_applications}
derives the relationship for root-mean-square speed versus temperature 
in the double truncated MB.
Finally, 
Section \ref{secjeans}
derives a new formula for Jeans flux in the exosphere. 

\section{The Maxwell-Boltzmann distribution}

\label{section_mb}
\label{mbsec}
Let $V$ be a random variable
defined in
$[0, \infty]$;
the MB 
probability density function (PDF), $f(v;a)$,
is
\begin{equation}
f (v;a) =
\frac
{
\sqrt {2}{v}^{2}{{\rm e}^{-\frac{1}{2}\,{\frac {{v}^{2}}{{a}^{2}}}}}
}
{
\sqrt {\pi}{a}^{3}
}
\quad ,
\end{equation}
where $a$ is a parameter and $v$ denotes the velocity,
see \cite{Maxwell1860,Boltzmann1872}.
Conversion to the physics is done by  
introducing the variable $a$, which is defined as  
\begin{equation}
a=\sqrt {{\frac {kT}{m}}}
\quad ,
\label{aphysics}
\end{equation}
where $m$ is  the mass of the gas molecules,
$k$ is the Boltzmann constant and 
$T$ is the thermodynamic temperature.
With this change of variable, the MB PDF
is 
\begin{equation}
f_p (v;m,k,T) =
\frac
{
\sqrt {2}{v}^{2}{{\rm e}^{-\frac{1}{2}\,{\frac {{v}^{2}m}{kT}}}}
}
{
\sqrt {\pi} \left( {\frac {kT}{m}} \right) ^{\frac{3}{2}}
}
\quad ,
\end{equation}
where the index $p$ stands for  physics.
The  distribution function (DF), $F(x;a)$,
is
\begin{subequations}
\begin{align}
F (v;a) =
\frac
{
\sqrt {2}{a}^{2} \left( a\sqrt {\pi}\sqrt {2}{\rm erf} \left(\frac{1}{2}\,{
\frac {\sqrt {2}v}{a}}\right)-2\,v{{\rm e}^{-\frac{1}{2}\,{\frac {{v}^{2}}{{a}
^{2}}}}} \right)
}
{
2\,\sqrt {\pi}{a}^{3}
}
\label{dfa}
\\
F_p(v) =  
\frac
{
\sqrt {2} \left(  \left( {\frac {kT}{m}} \right) ^{\frac{3}{2}}\sqrt {\pi}
\sqrt {2}{\rm erf} \left(\frac{1}{2}\,{\sqrt {2}v{\frac {1}{\sqrt {{\frac {kT
}{m}}}}}}\right)m-2\,v{{\rm e}^{-\frac{1}{2}\,{\frac {{v}^{2}m}{kT}}}}kT
 \right) 
}
{
2\,\sqrt {\pi} \left( {\frac {kT}{m}} \right) ^{\frac{3}{2}}m
}
\label{dfb}
\quad .
\end{align}
\end{subequations}
The average  value or mean, $\mu$,  is
\begin{subequations}
\begin{align}
\mu(a) =
2\,{\frac {\sqrt {2}a}{\sqrt {\pi}}}
\quad  , 
\\
\mu(m,k,T)_p= 
2\,{\frac {\sqrt {2}}{\sqrt {\pi}}\sqrt {{\frac {kT}{m}}}}
\quad ,
\end{align} 
\end{subequations}
the variance, $\sigma^2$, is
\begin{subequations}
\begin{align}
\sigma^2(a)=
{\frac {{a}^{2} \left( -8+3\,\pi \right) }{\pi}}
\\
\sigma^2 (m,k,T)_p=
{\frac {kT \left( -8+3\,\pi \right) }{m\pi}}
\quad .
\end{align}
\end{subequations}
The rth moment about the origin for the MB distribution is,
$\mu^{\prime}_r$,
is
\begin{subequations}
\begin{align}
\mu^{\prime}_r(a)=
\frac
{
{2}^{r/2+1}{a}^{r}\Gamma \left( r/2+\frac{3}{2} \right) 
}
{
\sqrt {\pi}
}
\\
\mu^{\prime}_r(m,k,T)_p=
\frac
{
{2}^{r/2+1} \left( \sqrt {{\frac {kT}{m}}} \right) ^{r}\Gamma \left( r
/2+\frac{3}{2} \right) 
}
{
\sqrt {\pi}
}
\quad ,
\end{align}
\end{subequations}
where
\begin{equation}
\mathop{\Gamma\/}\nolimits\!\left(z\right)
=\int_{0}^{\infty}e^{{-t}}t^{{z-1}}dt
\quad ,
\end{equation}
is the gamma function, see \cite{NIST2010}.
The root-mean-square speed , $v_{rms}$,
can be obtained from this formula by inserting $r=2$
\begin{subequations}
\begin{align}
v_{rms}(a) = \sqrt {3}a
\\
v_{rms}(m,k,T)_p =
\sqrt {3}\sqrt {{\frac {kT}{m}}}
\quad ,
\end{align}
\end {subequations}
see eqn.(7-10-16) in  \cite{Reif2009}.
This equation  allows us to derive the temperature 
once the root-mean-square speed  is measured
\begin{equation}
T = \frac{1}{3}\,{\frac {{v_{{{\it rms}}}}^{2}m}{k}}
\quad  .
\label{tfromrms}
\end{equation}
The coefficient of variation (CV) is
\begin{equation}
CV =\frac{\sigma(a)}{\mu(a)}
=\sqrt {\frac{3}{8}\,\pi-1}
\quad ,
\end{equation}
which is constant. 
The first three rth moments about the mean for the MB distribution,
$\mu_r(a)$,  are  
\begin{subequations}
\begin{align}
\mu_2(a)={\frac {{a}^{2} \left( -8+3\,\pi \right) }{\pi}}
\\
\mu_3(a)=-2\,{\frac {{a}^{3}\sqrt {2} \left( 5\,\pi-16 \right)
}{{\pi}^{3/2}}}
\\
\mu_4(a)={\frac {{a}^{4} \left( 15\,{\pi}^{2}+16\,\pi-192 \right) }{{\pi}^{2}}}
\quad  .
\end{align}
\end{subequations}

The mode is at 
\begin{subequations}
\begin{align}
v(a)=\sqrt {2}a \\
\label{modekt}
v(m,k,T)_p= \sqrt {2}\sqrt {{\frac {kT}{m}}}
\quad .
\end{align}
\end{subequations}
An approximate expression for the median can be
obtained by a Taylor series of the DF around the mode.
The approximation formula is
\begin{subequations}
\begin{align}
v(a)=-\frac{1}{4}\,a \left( -6+{\rm e} \left( {\rm erf} \left(1\right)-\frac{1}{2} \right) 
\sqrt {\pi} \right) \sqrt {2}
\quad ,
\\
v(m,k,T)_p=
-\frac{1}{4}\,\sqrt {{\frac {kT}{m}}} \left( -6+{\rm e} \left( {\rm erf} 
\left(1\right)-\frac{1}{2} \right) \sqrt {\pi} \right) \sqrt {2}
\quad ,
\end{align}
\end{subequations}
which has a percent error, $\delta$, of $\delta \approx 0.04 \%$
in respect to the numerical value.
The entropy is
\begin{subequations}
\begin{align}
\ln  \left( \sqrt {2}\sqrt {\pi}a \right) -\frac{1}{2}+\gamma
\quad ,
\\
\ln  \left( \sqrt {2}\sqrt {\pi}\sqrt {{\frac {kT}{m}}} \right) 
-\frac{1}{2}+
\gamma
\quad ,
\end{align}
\end{subequations}
where $\gamma$ is the Euler-Mascheroni constant, which is
defined as
\begin{equation}
\gamma=\lim_{n\to\infty}\left(1+\frac{1}{2}+\frac{1}{3}+\dots+\frac{1}{n}-\ln
n \right)=0.57721\,\dots.
\quad ,
\end{equation}
see \cite{NIST2010} for more details.
The coefficient of skewness is 
\begin{equation}
{\frac { \left( -10\,\pi+32 \right) \sqrt {2}}{ \left( -8+3\,\pi
 \right) ^{\frac{3}{2}}}} \approx 0.48569
\quad ,
\end{equation}
and
the coefficient of kurtosis is
\begin{equation}
{\frac {15\,{\pi}^{2}+16\,\pi-192}{ \left( -8+3\,\pi \right) ^{2}}}
\approx 3.10816
\quad .
\end{equation}

According to \cite{Devroye1986},  
a random number generation can be  obtained
via  inverse transform sampling when
the distribution function or cumulative distribution 
function, $F(x)$, is known:
(i) a pseudo  number generator gives  
      a random  number ${\bf R}$ between zero and one; 
(ii)  the inverse function x= $F^{-1}(\bf{R})$ is evaluated; and
(iii) the procedure is repeated for different  values of $\bf{R}$. 
In our case, the inverse function 
should be evaluated in  a numerical way    
by solving for $v$ the following nonlinear equation
\begin{subequations}
\begin{align}
F(v;a) - {\bf R} =0
\label{eqnarandom}
\quad ,
\\
F(v;m,k,T)_p - {\bf R} =0
\quad ,
\end{align}
\end{subequations}
where $F(v)$  and  $F_p(v)$ are the two DF
represented by equations (\ref{dfa}) and (\ref{dfb}).
As a practical example, by
inserting in equation (\ref{eqnarandom}) $a=1$ 
and $\bf{R}=0.5$, 
we obtain in a numerical way $v=1.538$.

\section{The double truncated Maxwell-Boltzmann distribution}

\label{section_mb_truncated}
Let $V$ be a random variable that is
defined in
$[v_l,v_u]$;
the {\em double truncated } 
version of the Maxwell-Boltzmann PDF, 
 $f_t(v;a,v_l,v_u)$,
is
\begin{equation}
f_t(v;a,v_l,v_u) =
{v}^{2}{{\rm e}^{-\frac{1}{2}\,{\frac {{v}^{2}}{{a}^{2}}}}}
\quad ,
\end{equation}
where 
\begin{equation}
C= \frac
{
-2
}
{
CD
}
\quad ,
\end{equation}
where 
\begin{eqnarray}
CD =
{a}^{2} \Bigg( -a\sqrt {\pi}\sqrt {2}{\rm erf} \left(\frac{1}{2}\,{\frac {
\sqrt {2}{\it vu}}{a}}\right)+a\sqrt {\pi}\sqrt {2}{\rm erf} \left(\frac{1}{2}
\,{\frac {\sqrt {2}{\it vl}}{a}}\right)
\nonumber \\
+2\,{\it vu}\,{{\rm e}^{-\frac{1}{2}\,{
\frac {{{\it vu}}^{2}}{{a}^{2}}}}}-2\,{\it vl}\,{{\rm e}^{-\frac{1}{2}\,{
\frac {{{\it vl}}^{2}}{{a}^{2}}}}} \Bigg) 
\quad ,
\end{eqnarray}
and  
 ${\rm erf(x)}$ is the error function, which is defined as
\begin{equation}
\mathop{\mathrm{erf}\/}\nolimits
(x)=\frac{2}{\sqrt{\pi}}\int_{0}^{x}e^{-t^{2}}dt
\quad ,
\end{equation}
see \cite{NIST2010}.
The physical meaning of $a$ is still represented by 
equation (\ref{aphysics}); however, due to the tendency 
to obtain complicated expressions, we will omit the
double notation.  
The  DF, $F_t(v;a,v_l,v_u)$,
is
\begin{equation}
F_t(v;a,v_l,v_u)
=
\frac
{
C{a}^{2} \left( \sqrt {\pi}\sqrt {2}a\,{\rm erf} \left(\frac{1}{2}\,{\frac {
\sqrt {2}v}{a}}\right)-2\,v{{\rm e}^{-\frac{1}{2}\,{\frac {{v}^{2}}{{a}^{2}}}}
} \right) 
}
{
2
}
\quad .
\end{equation}
The average  value  $\mu_t(a,v_l,v_u)$,  is
\begin{eqnarray}
\mu_t(a,v_l,v_u
=
C{a}^{2} \Bigg( 2\,{{\rm e}^{-\frac{1}{2}\,{\frac {{v_{{l}}}^{2}}{{a}^{2}}}}}{
a}^{2}-2\,{{\rm e}^{-\frac{1}{2}\,{\frac {{v_{{u}}}^{2}}{{a}^{2}}}}}{a}^{2}+{
{\rm e}^{-\frac{1}{2}\,{\frac {{v_{{l}}}^{2}}{{a}^{2}}}}}{v_{{l}}}^{2}
\nonumber \\
-{
{\rm e}^{-\frac{1}{2}\,{\frac {{v_{{u}}}^{2}}{{a}^{2}}}}}{v_{{u}}}^{2}
 \Bigg) 
\quad. 
\end{eqnarray}
The rth moment about the origin for the double truncated 
MB distribution is,
$\mu^{\prime}_{r,t}(a,v_l,v_u)$,
\begin{equation}
\mu^{\prime}_{r,t}(a,v_l,v_u)
=
\frac
{MN}
{r+3}
\quad 
\end{equation}
where 
\begin{eqnarray}
MN=
C{2}^{\frac{r}{4}+\frac{5}{4}}{a}^{2} \times 
\nonumber \\
 \Bigg (  \left( {\frac {{v_{{u}}}^{2}}{{a}^{2}}}
 \right) ^{-\frac{r}{4}-\frac{1}{4}}{v_{{u}}}^{r+1}{{\rm e}^{-\frac{1}{4}\,{\frac {{v_{{u}}}^{
2}}{{a}^{2}}}}}{{\sl M}_{\frac{r}{4}+\frac{1}{4},\,\frac{r}{4}+\frac{3}{4}}\left(\frac{1}{2}\,{\frac {{v_{{u}}
}^{2}}{{a}^{2}}}\right)}
\nonumber \\
-{v_{{l}}}^{r+1}{{\rm e}^{-\frac{1}{4}\,{\frac {{v_{{l
}}}^{2}}{{a}^{2}}}}}{{\sl M}_{\frac{r}{4}+\frac{1}{4},\,\frac{r}{4}+\frac{3}{4}}\left(\frac{1}{2}\,{\frac {{v_
{{l}}}^{2}}{{a}^{2}}}\right)} \left( {\frac {{v_{{l}}}^{2}}{{a}^{2}}}
 \right) ^{-\frac{r}{4}-\frac{1}{4}} \Bigg) 
\end{eqnarray}
where 
${{\sl M}_{\mu,\,\nu}\left(z\right)}$ is the 
Whittaker M function, see \cite{NIST2010}.
The root-mean-square speed, $v_{rms,t}(a,v_l,v_u) $,
can be obtained from this formula by inserting $r=2$,
and is 
\begin{equation}
(v_{rms,t} (a,v_l,v_u)) = \sqrt{\frac{NV}
{
5\, \left( {\frac {{v_{{u}}}^{2}}{{a}^{2}}} \right) ^{3/4} \left( {
\frac {{v_{{l}}}^{2}}{{a}^{2}}} \right) ^{3/4}
}} 
\quad ,
\label{vrmstrunc}
\end {equation}
where 
\begin{eqnarray}
NV=
2\,C{2}^{3/4}{a}^{2} \Bigg ( {v_{{u}}}^{3}{{\rm e}^{-1/4\,{\frac {{v_{{
u}}}^{2}}{{a}^{2}}}}}{{\sl M}_{3/4,\,5/4}\left(1/2\,{\frac {{v_{{u}}}^
{2}}{{a}^{2}}}\right)} \left( {\frac {{v_{{l}}}^{2}}{{a}^{2}}}
 \right) ^{3/4}
\nonumber \\
-{v_{{l}}}^{3}{{\rm e}^{-1/4\,{\frac {{v_{{l}}}^{2}}{{a
}^{2}}}}}{{\sl M}_{3/4,\,5/4}\left(1/2\,{\frac {{v_{{l}}}^{2}}{{a}^{2}
}}\right)} \left( {\frac {{v_{{u}}}^{2}}{{a}^{2}}} \right) ^{3/4}
 \Bigg) 
\quad  .
\end{eqnarray}

The variance $\sigma^2_t(a,v_l,v_u) $ is defined as 
\begin{equation}
\sigma^2_t(a,v_l,v_u) 
=
\mu^{\prime}_{2,t}(a,v_l,v_u)- \big (\mu^{\prime}_{1,t}(a,v_l,v_u)\big)^2
\end{equation}
and  has the following explicit form
\begin{eqnarray}
\sigma^2_t(a,v_l,v_u)
= 
\nonumber \\
4\, \Bigg ( \Big   (    ( v_{{l}}+2\,v_{{u}}   ) {a}^{2}+v_{{l}}v
_{{u}}   ( v_{{l}}+\frac{1}{2}\,v_{{u}}   ) \Big   )    ( {a}^{2}+1/
2\,{v_{{u}}}^{2}   ) {C}^{2}{a}^{4}{{\rm e}^{-\frac{1}{2}\,{\frac {{v_{{l}
}}^{2}+2\,{v_{{u}}}^{2}}{{a}^{2}}}}}
\nonumber \\
-2\, \Big  (    ( v_{{l}}+\frac{1}{2}\,v
_{{u}}   ) {a}^{2}
+\frac{1}{4}\,v_{{l}}v_{{u}}   ( v_{{l}}+2\,v_{{u}}
   )  \Big  ) {C}^{2}{a}^{4}   ( {a}^{2}+\frac{1}{2}\,{v_{{l}}}^{2}
   ) {{\rm e}^{-\frac{1}{2}\,{\frac {2\,{v_{{l}}}^{2}+{v_{{u}}}^{2}}{{a}^{
2}}}}}
\nonumber \\
+   ( {a}^{2}+\frac{1}{2}\,{v_{{u}}}^{2}   ) \Big    ( C{\rm \, erf} 
  (\frac{1}{2}\,{\frac {\sqrt {2}v_{{l}}}{a}}  ){a}^{3}\sqrt {2}\sqrt {
\pi}
\nonumber \\
-C{\rm \, erf}   (\frac{1}{2}\,{\frac {\sqrt {2}v_{{u}}}{a}}  ){a}^{3}
\sqrt {2}\sqrt {\pi}+4 \Big  ) C{a}^{2}   ( {a}^{2}+\frac{1}{2}\,{v_{{l}}}^
{2}   ) {{\rm e}^{-\frac{1}{2}\,{\frac {{v_{{l}}}^{2}+{v_{{u}}}^{2}}{{a}^{
2}}}}}
\nonumber \\
+{C}^{2}{a}^{4}   ( {a}^{2}+\frac{1}{2}\,{v_{{l}}}^{2}   ) ^{2}v_
{{l}}{{\rm e}^{-\frac{3}{2}\,{\frac {{v_{{l}}}^{2}}{{a}^{2}}}}}-   ( {a}^{2
}+\frac{1}{2}\,{v_{{u}}}^{2}   ) ^{2}{C}^{2}{a}^{4}v_{{u}}{{\rm e}^{-\frac{3}{2}\,
{\frac {{v_{{u}}}^{2}}{{a}^{2}}}}}
\nonumber \\
+   ( \frac{3}{4}\,{a}^{2}v_{{l}}+\frac{1}{4}\,{v
_{{l}}}^{3}   ) {{\rm e}^{-\frac{1}{2}\,{\frac {{v_{{l}}}^{2}}{{a}^{2}}}}}
+   ( -\frac{3}{4}\,{a}^{2}v_{{u}}-\frac{1}{4}\,{v_{{u}}}^{3}   ) {{\rm e}^{-1/
2\,{\frac {{v_{{u}}}^{2}}{{a}^{2}}}}}
\nonumber \\
-\frac{1}{2}\,  \Big ( \Big   ( C{\rm \, erf} 
  (\frac{1}{2}\,{\frac {\sqrt {2}v_{{l}}}{a}}  ){a}^{3}\sqrt {2}\sqrt {
\pi}
\nonumber \\
-C{\rm \, erf}   (\frac{1}{2}\,{\frac {\sqrt {2}v_{{u}}}{a}}  ){a}^{3}
\sqrt {2}\sqrt {\pi}+4 \Big  ) C   ( {a}^{2}+\frac{1}{2}\,{v_{{l}}}^{2}
   ) ^{2}{{\rm e}^{-{\frac {{v_{{l}}}^{2}}{{a}^{2}}}}}
\nonumber \\
+   ( {a}
^{2}+\frac{1}{2}\,{v_{{u}}}^{2}   ) ^{2} \Big  ( C{\rm \, erf}   (\frac{1}{2}\,{
\frac {\sqrt {2}v_{{l}}}{a}}  ){a}^{3}\sqrt {2}\sqrt {\pi}
\nonumber \\
-C
{\rm \, erf}   (\frac{1}{2}\,{\frac {\sqrt {2}v_{{u}}}{a}}  ){a}^{3}\sqrt 
{2}\sqrt {\pi}+4  \Big ) C{{\rm e}^{-{\frac {{v_{{u}}}^{2}}{{a}^{2}}}}
}
+\frac{3}{4}\,\sqrt {\pi}\sqrt {2}  \Big ( -{\rm \, erf}   (\frac{1}{2}\,{\frac {
\sqrt {2}v_{{u}}}{a}}  )
\nonumber \\
+{\rm \, erf}   (\frac{1}{2}\,{\frac {\sqrt {2}v_{
{l}}}{a}}  )  \Big ) a \Big  ) {a}^{2} \Bigg) C{a}^{2}
\quad  .
\end{eqnarray}
Although the coefficients of skewness and kurtosis  
for the truncated MB exist, they have a complicated expression.

\section{A laboratory application}

\label{section_mb_applications}
\begin{figure*}
\begin{center}
\includegraphics[width=5cm]{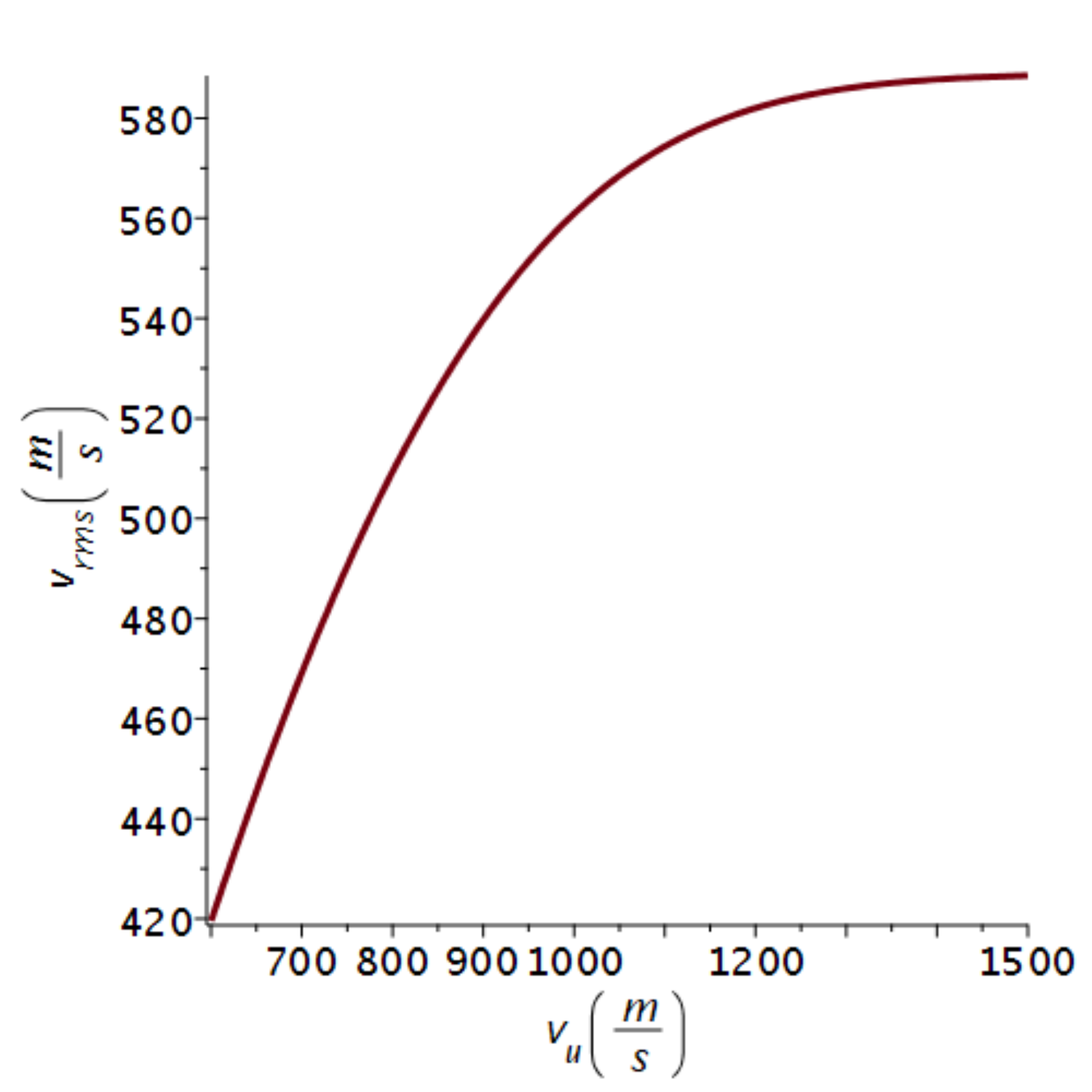}
\end {center}
\caption
{
The theoretical root-mean-square speed  as a function of the upper limit 
in velocity (continuous line) 
and standard value of the temperature (dotted line)
when $a$=340 and $v_l=0$.
}
\label{vrmsvu}
    \end{figure*}

\begin{figure*}
\begin{center}
\includegraphics[width=5cm,angle=-90]{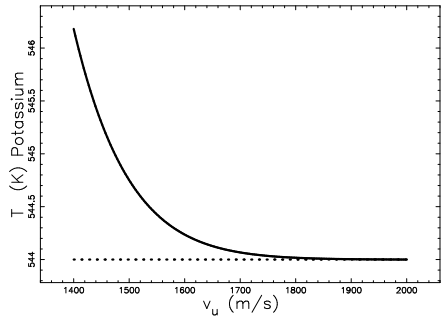}
\end {center}
\caption
{
Temperature as a function of the upper limit 
in velocity for Potassium (continuous line) 
and standard value of the temperature (dotted line)
when $v_l=$ and  $v_{rms,m}=589.111511 m/s $
}
\label{temp_potassium}
    \end{figure*}

The temperature  as a function of 
root-mean-square speed  for the MB  
is given by equation (\ref{tfromrms}).
In the truncated  MB distribution, the temperature  can be  found
by solving the following   nonlinear equation
\begin{equation}
v_{rms,t} (k,m,T,v_l,v_u)=v_{rms,m}
\quad ,
\label{eqnnlvrmst}
\end{equation}
where  $v_{rms,m}$ 
is not a theoretical  variable 
but is 
the root-mean-square speed  measured 
in the laboratory
and $v_{rms,t}$ is given by equation (\ref{vrmstrunc}).
The  laboratory measures of
$v_{rms,m}$ 
started with \cite {Eldridge1927},
where a  $v_{rms,m}$=388 m/s  at $400~^{\circ}C$ was 
found for a metallic vapor.
In the truncated MB distribution, there are three parameters that 
can be measured in the laboratory from a kinematical point 
of view, as follows: 
the lowest  velocity, $v_l$;
the highest velocity, $v_u$; and
the root-mean-square speed, $v_{rms,m}$.
Setting for simplicity $v_l$=0, we will now 
explore the effect of 
the variation of $v_u$ on the root-mean-square speed;
see Figure \ref{vrmsvu}.
The {\it first} example 
of the influence of the upper limit 
in velocity on the temperature 
is 
 given by  potassium gas 
\cite{Miller1955,Hernandez2017},
in which molecular mass is $ 6.492429890\,10^{-26}\,kg$.
In Figure \ref{temp_potassium}, we evaluate 
in a numerical way 
the temperature when $v_l$=0 and 
$v_u$ is variable in the case of a measured value of $v_{rms,m}$.

\begin{figure*}
\begin{center}
\includegraphics[width=5cm,angle=-90]{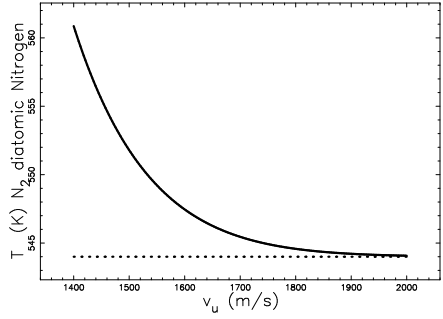}
\end {center}
\caption
{
Temperature as a function of the upper limit 
in velocity for diatomic nitrogen, $N_2$, (continuous line) 
and standard value of the temperature (dotted line)
when $v_l=$ and  $v_{rms,m}=695.9756308 m/s $
}
\label{temp_nitrogen2}
    \end{figure*}

The {\it second} example is given by  diatomic nitrogen, $N_2$,  
in which molecular mass is $ 4.651737684\,10^{-26}\,kg$.
In Figure \ref{temp_nitrogen2}, we evaluate 
the temperature when $v_l$=0 and 
$v_u$ is a variable in the case of a measured value of $v_{rms,m}$.

\section{The Jeans escape}

The standard formula  for the escape of molecules 
from the exosphere is reviewed  in the framework 
of the MB distribution.
A new formula for the Jeans escape  is derived
in the
framework of the truncated MB.

\subsection{The standard case}

\label{secfluxstandard}
In the exosphere,    a molecule of mass  $m$ and
velocity $v_e$ is free to escape when
\begin{equation}
\frac{1}{2} m v_e^2 - G \frac{M m}{R_{ex}} =0
\quad  ,
\end {equation}
where 
$G$ is the Newtonian gravitational constant,
$M$ is the mass of the Earth,
$R_{ex}=R+H $ is the radius of the exosphere,
$R$ is the radius of the Earth and $H$ is the altitude 
of the exosphere.
The flux of the molecules that are living in the exosphere 
$\Phi_j$ 
is 
\begin{equation} 
\Phi_j= \frac{1}{4} N_{ex} \mu_e 
\quad , 
\end{equation}
where 
$N_{ex}$ is the
number  of molecules per unit volume
and $\mu_e$ is the average velocity  of escape.
In the  presence of a given number  of molecules per unit volume, 
the standard  MB distribution in velocities in a unit volume, $f_m$,
is 
\begin{equation}
f_m (v;m,k,T,N_{ex}) = N_{ex} 
\frac
{
\sqrt {2}{v}^{2}{{\rm e}^{-\frac{1}{2}\,{\frac {{v}^{2}m}{kT}}}}
}
{
\sqrt {\pi} \left( {\frac {kT}{m}} \right) ^{\frac{3}{2}}
}
\quad .
\end{equation}
The average value of escape is defined as 
\begin{equation}
\mu_e =
\frac
{
\int_{v_e}^\infty v f_m (v;m,k,T,N_{ex}) dv
}
{
\int_{0}^\infty  f_m (v;m,k,T,N_{ex}) dv
}
\quad .
\end{equation}
In this integral, the following changes are made to the variables 
\begin{equation}
\lambda = \frac{1}{2} \frac{mv^2}{kT}
\quad .
\label{eqnlambda}
\end{equation}
Therefore, 
\begin{equation}
\mu_e = 2\, \left( \lambda_{{e}}+1 \right) {{\rm e}^{-\lambda_{{e}}}}\sqrt
{2}
\sqrt {{\frac {kT}{\pi\,m}}}
\quad ,
\end{equation}
with 
\begin{equation}
\lambda_e= 2\,{\frac {GM}{R_{{{\it ex}}}{v_{{0}}}^{2}}}
\quad ,
\end{equation}
where $v_0$ is the mode as represented 
by equation (\ref{modekt}).
The flux is now
\begin{equation} 
\Phi_j=
\frac
{
N_{{{\it ex}}} \left( \lambda_{{e}}+1 \right) {{\rm e}^{-\lambda_{{e}}
}}v_{{0}}
}
{
2\,\sqrt {\pi}
}
\quad .
\end{equation}
For more details see \cite{Jeans1955,Shu1982,Catling2017,Owen2019}.
On adopting the parameters of Table \ref{tableexo}
the Jeans escape flux for hydrogen is
\begin{equation} 
\Phi_j= 3.98 \,10^{11} molecules \, m^{-2} s^{-1}
\quad , 
\end{equation}
and  
\begin{equation}
\lambda_e = 7.78
\quad .
\end{equation}

The Jeans escape flux for Earth 
at $T=900\,K$ varies between   
$\Phi_j \approx 2.7  \,10^{11} molecules \, m^{-2} s^{-1}$; 
see \cite{Vidal-Madjar1974} or  
 Fig.1 in  \cite{Liu2017}.
and
$\Phi_j \approx 4  \,10^{11} molecules \, m^{-2} s^{-1}$, 
see \cite{Bertaux1974}.
Therefore, our choice of parameters is compatible
with the suggested interval in  flux. 

\begin{table}[ht!]
\caption
{
Adopted physical parameters for the exosphere
}
\label{tableexo}
\begin{tabular}{|c|c|}
\hline
Parameter & value  \\
\hline
$R_{ex}$  & 6900 km   \\
T         & 900  K    \\
$N_{ex}$  & $10^{11}m^{-3}$ \\ 
\hline
\end{tabular}
\end{table}

\subsection{The truncated case}
\label{secjeans}
\begin{figure*}
\begin{center}
\includegraphics[width=5cm]{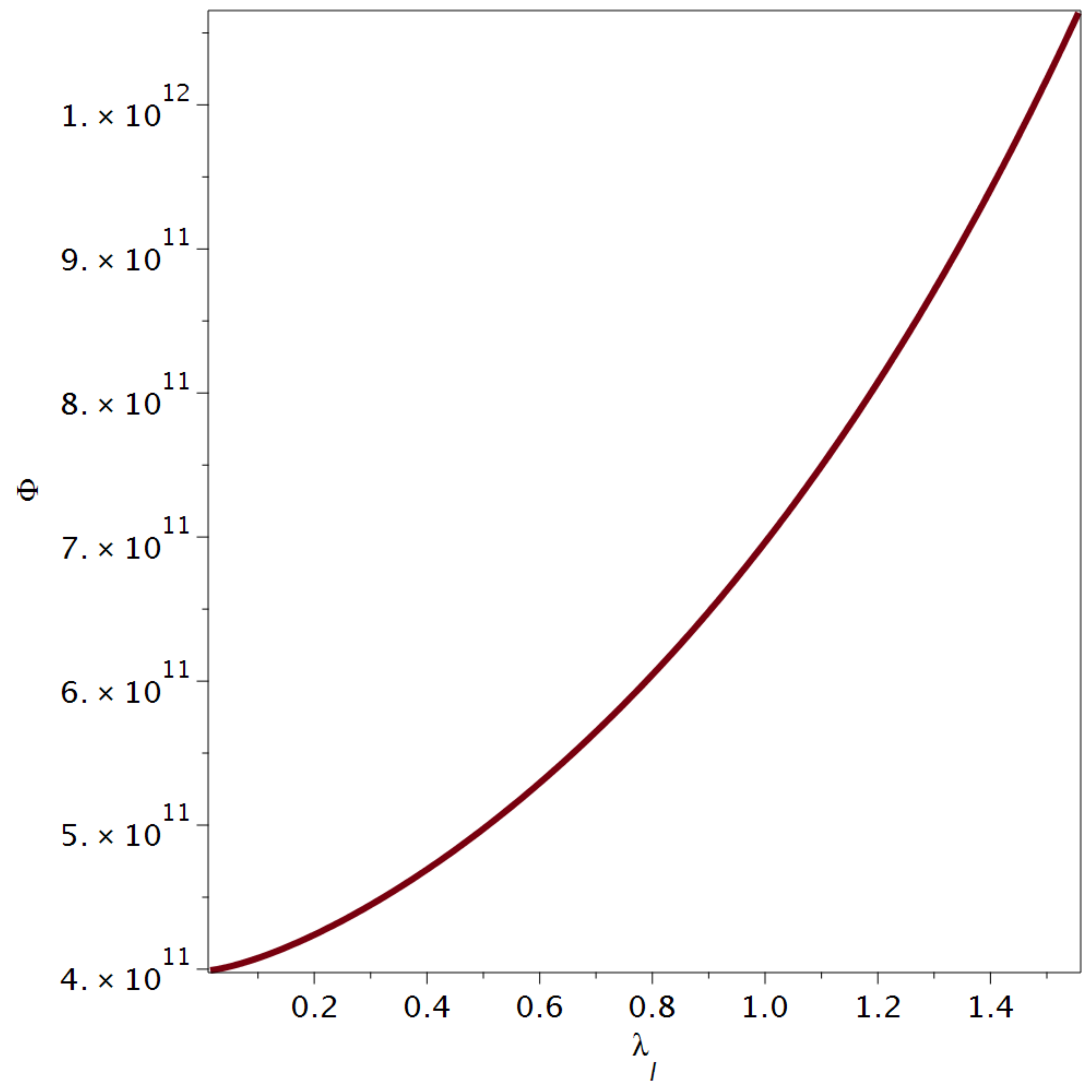}
\end {center}
\caption
{
The flux of molecules as a function of 
$\lambda_l$
with  parameters as in Table \ref{tableexo},
$\lambda_e = 7.78$ 
and  $\lambda_u=1000 \lambda_e$.
}
\label{fluxtrunc}
    \end{figure*}

\begin{figure*}
\begin{center}
\includegraphics[width=8cm]{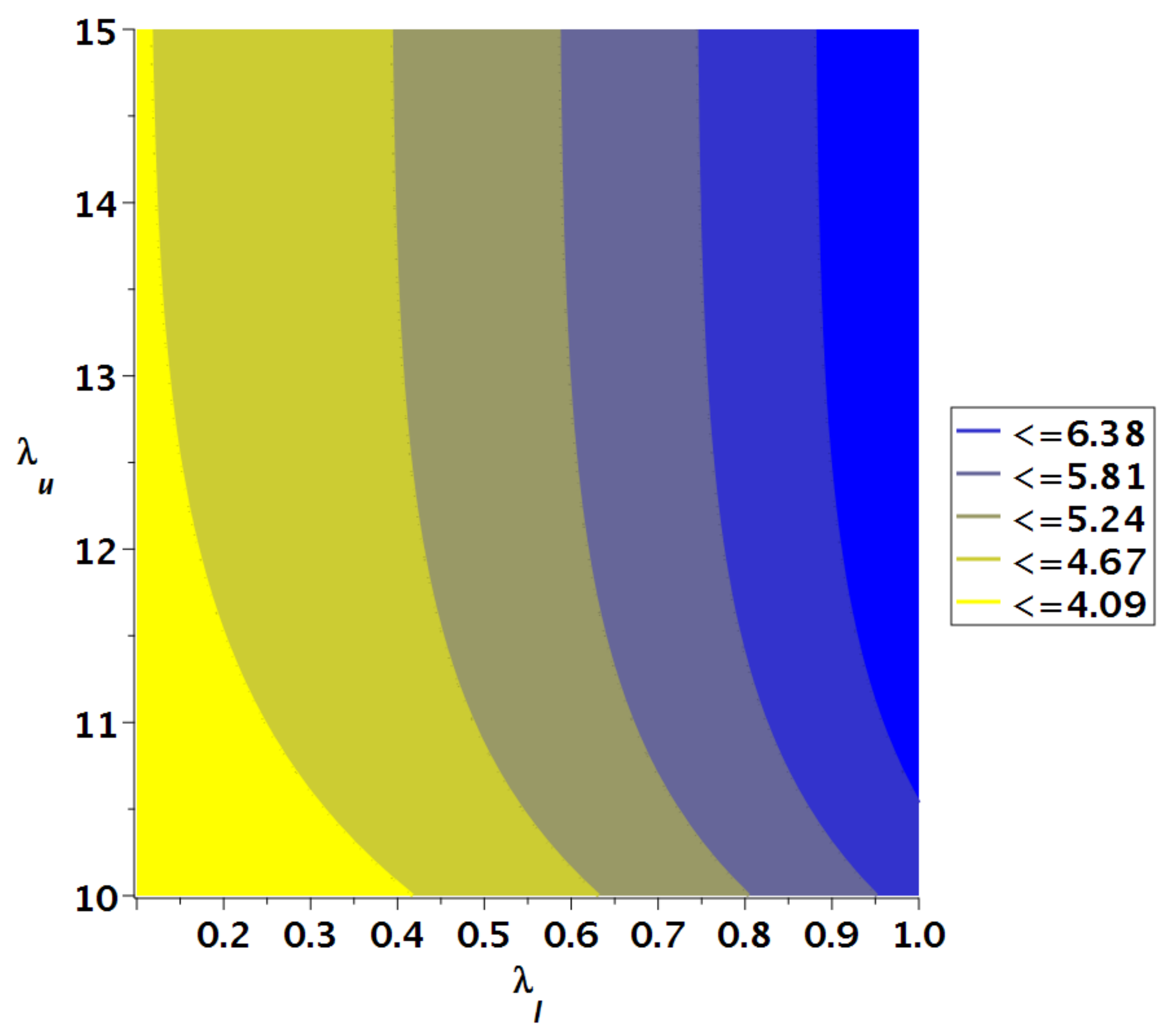}
\end {center}
\caption
{
The flux of molecules as a function of 
$\lambda_l$ and $\lambda_u$
with  parameters as in 
Table \ref{tableexo}.
}
\label{matrixtrunc}
    \end{figure*}

The average value of escape 
for a truncated MB distribution, $\mu_{e,t}$,
is 
\begin{equation}
\mu_{e,t} =
\frac
{
\int_{v_e}^\infty v f_t (v;m,k,T,N_{ex},v_l,v_u) dv
}
{
\int_{0}^\infty  f_m (v;m,k,T,N_{ex},v_l,v_u) dv
}
\quad .
\end{equation}
This integral can be solved 
by introducing the change of variable as given 
by equation (\ref{eqnlambda}) 
\begin{equation}
\mu_{e,t} =
-2\,{\frac { \left(  \left( \lambda_{{u}}+1 \right) {{\rm e}^{-\lambda
_{{u}}}}-{{\rm e}^{-\lambda_{{e}}}} \left( \lambda_{{e}}+1 \right) 
 \right) \sqrt {2}}{2\,\sqrt {\lambda_{{l}}}{{\rm e}^{-\lambda_{{l}}}}
-2\,\sqrt {\lambda_{{u}}}{{\rm e}^{-\lambda_{{u}}}}-\sqrt {\pi}
{\rm erf} \left(\sqrt {\lambda_{{l}}}\right)+\sqrt {\pi}{\rm erf} 
\left(\sqrt {\lambda_{{u}}}\right)}\sqrt {{\frac {kT}{m}}}}
\quad  ,
\end{equation}
where 
$\lambda_l$ is the lower value of $\lambda$ 
and
$\lambda_u$ is the upper value of $\lambda$.
The flux of the molecules that are living the exosphere 
in the truncated MB,
$\Phi_{j,t}$, is 
\begin{equation}
\Phi_{j,t} =
{\frac {N_{{{\it ex}}} \left(  \left( \lambda_{{u}}+1 \right) {{\rm e}
^{-\lambda_{{u}}}}-{{\rm e}^{-\lambda_{{e}}}} \left( \lambda_{{e}}+1
 \right)  \right) \sqrt {2}}{4\,\sqrt {\lambda_{{u}}}{{\rm e}^{-
\lambda_{{u}}}}+2\,\sqrt {\pi}{\rm erf} \left(\sqrt {\lambda_{{l}}}
\right)-2\,\sqrt {\pi}{\rm erf} \left(\sqrt {\lambda_{{u}}}\right)-4\,
\sqrt {\lambda_{{l}}}{{\rm e}^{-\lambda_{{l}}}}}\sqrt {{\frac {kT}{m}}
}}
\quad  .
\label{jeanstruncated}
\end{equation}
The  increasing   flux of molecules
is outlined when one parameter, $\lambda_l$,
is variable; 
see Figure \ref{fluxtrunc}.
In other words, an increase in $\lambda_l$ produces an increase 
in the flux of the molecules.
The dependence of the flux
when two parameters 
are variable, $\lambda_l$ and $\lambda_u$,
is reported in  Figure \ref{matrixtrunc}.

These Jeans escape fluxes  for Earth 
are  compatible   with the observed values
that were reported in Section \ref{secfluxstandard}.

\section{Conclusions}

This paper derived analytical formulae for the following 
quantities for  a double  truncated MB distribution:
the PDF, 
the DF, 
the average value,
the rth moment about the origin,
the root-mean-square speed
and the variance.
The traditional correspondence between 
root-mean-square speed and temperature is 
replaced by the nonlinear Equation(\ref{eqnnlvrmst}).
The   new formula  (\ref{jeanstruncated}) 
for  the Jeans escape flux
of molecules from an atmosphere is now a function 
of the lower and upper boundary in velocity.  


\begin{thebibliography}{10}
\expandafter\ifx\csname url\endcsname\relax
  \def\url#1{{\tt #1}}\fi
\expandafter\ifx\csname urlprefix\endcsname\relax\def\urlprefix{URL }\fi
\providecommand{\eprint}[2][]{\url{#2}}

\bibitem{Maxwell1860}
Maxwell J~C 1860 V. Illustrations of the dynamical theory of gases.—part i.
  on the motions and collisions of perfectly elastic spheres {\em The London,
  Edinburgh, and Dublin Philosophical Magazine and Journal of Science\/} {\bf
  19}(124), 19

\bibitem{Boltzmann1872}
{Boltzmann} L 1872 Weitere studien über das wärmegleichgewicht unter
  gasmolekülen {\em K. Acad. Wiss.(Wein) Sitzb., II Abt\/} {\bf 66}, 275

\bibitem{Buzzi1970}
Buzzi J, Doucet H and Gresillon D 1970 Ion distribution functions in
  collisionless surface ionized plasmas {\em The Physics of Fluids\/} {\bf
  13}(12), 3041

\bibitem{Treguier1976}
Treguier J and Henry D 1976 Propagation of electronic longitudinal modes in a
  truncated Maxwellian plasma {\em Journal of Plasma Physics\/} {\bf 15}(3),
  447

\bibitem{Kishimoto1983}
Kishimoto Y, Mima K, Watanabe T and Nishikawa K 1983 Analysis of fast-ion
  velocity distributions in laser plasmas with a truncated Maxwellian velocity
  distribution of hot electrons {\em The Physics of Fluids\/} {\bf 26}(8), 2308

\bibitem{Tomita2007}
Tomita Y, Smirnov R, Nakamura H, Zhu S, Takizuka T and Tskhakaya D 2007 Effect
  of truncation of electron velocity distribution on release of dust particle
  from plasma-facing wall {\em Journal of Nuclear Materials\/} {\bf 363}, 264

\bibitem{Fowlie2017}
{Fowlie} A 2017 {Halo-independence with quantified maximum entropy at
  DAMA/LIBRA} {\em \jcap\/} {\bf 2017}(10) 002 (\textit{Preprint}
  \eprint{1708.00181})

\bibitem{Ida2017}
{Ida} K, {Kobayashi} T, {Yoshinuma} M, {Akiyama} T, {Tokuzawa} T, {Tsuchiya} H,
  {Itoh} K and {LHD Experiment Group} 2017 {Observation of distorted
  Maxwell-Boltzmann distribution of epithermal ions in LHD} {\em Physics of
  Plasmas\/} {\bf 24}(12) 122502

\bibitem{Todorov2019}
{Todorov} P, {de Aquino Carvalho} J~C, {Maurin} I, {Laliotis} A and {Bloch} D
  2019 {Search for deviations from the ideal Maxwell-Boltzmann distribution for
  a gas at an interface} in {\em \procspie\/} vol 11047 of {\em Society of
  Photo-Optical Instrumentation Engineers (SPIE) Conference Series\/} p 110470P
  (\textit{Preprint} \eprint{1810.04876})

\bibitem{NIST2010}
Olver F~W~J~e, Lozier D~W~e, Boisvert R~F~e and Clark C~W~e 2010 {\em {NIST
  {Handbook} of {Mathematical} {Functions}}\/} (Cambridge: {Cambridge
  University Press. })

\bibitem{Reif2009}
Reif F 2009 {\em Fundamentals of Statistical and Thermal Physics\/} (Long
  Grove, Illinois: Waveland Press) ISBN 9781478610052

\bibitem{Devroye1986}
Devroye L 1986 {\em General principles in random variate generation\/} (New
  York: Springer)

\bibitem{Eldridge1927}
Eldridge J 1927 Experimental test of Maxwell's distribution law {\em Physical
  Review\/} {\bf 30}(6), 931

\bibitem{Miller1955}
{Miller} R~C and {Kusch} P 1955 {Velocity distributions in potassium and
  thallium atomic beams} {\em Physical Review\/} {\bf 99}(4), 1314

\bibitem{Hernandez2017}
Hernandez H 2017 Standard Maxwell-Boltzmann distribution: Definition and
  properties {\em ForsChem Research Reports\/} {\bf 2}

\bibitem{Jeans1955}
{Jeans} J~H 1955 {\em {The dynamical theory of gases}\/} (New York: Dover)

\bibitem{Shu1982}
{Shu} F~H 1982 {\em {The Physical Universe}\/} (Mill Valley CA: University
  Science Books)

\bibitem{Catling2017}
Catling D and Kasting J 2017 Escape of atmospheres to space {\em Atmospheric
  Evolution on Inhabited and Lifeless Worlds (Cambridge, 2017) chap\/} {\bf 5},
  129

\bibitem{Owen2019}
{Owen} J~E 2019 {Atmospheric escape and the evolution of close-in exoplanets}
  {\em Annual Review of Earth and Planetary Sciences\/} {\bf 47}, 67
  (\textit{Preprint} \eprint{1807.07609})

\bibitem{Vidal-Madjar1974}
Vidal-Madjar A, Blamont J and Phissamay B 1974 Evolution with solar activity of
  the atomic hydrogen density at 100 kilometers of altitude {\em Journal of
  Geophysical Research (1896-1977)\/} {\bf 79}(1), 233

\bibitem{Liu2017}
{Liu} W, {Chiao} M, {Collier} M~R and et~al 2017 {The structure of the local
  hot bubble} {\em The Astrophysical Journal\/} {\bf 834}(1) 33
  (\textit{Preprint} \eprint{1611.05133})

\bibitem{Bertaux1974}
Bertaux J 1974 Lhydrog{\`e}ne atomique dans lexosph{\`e}re terrestre: mesures
  dintensit{\'e} et de largeur de raie de l{\'e}mission lyman alpha {\`a} bord
  du satellite ogo 5 et interpr{\'e}tation {\em These d{\'e}tat, Universit{\'e}
  Paris\/} {\bf 6}

\end{thebibliography}
\providecommand{\newblock}{}

\end{document}